\documentstyle{article}

\font\tenrm=cmr10
\font\tenit=cmti10
\font\elevenbf=cmbx10 scaled\magstep 1
\font\elevenrm=cmr10 scaled\magstep 1
 1

\font\ninerm=cmr9

\renewenvironment{thebibliography}[1]
 { \elevenrm
   \begin{list}{\arabic{enumi}.}
    {\usecounter{enumi} \setlength{\parsep}{0pt}
     \setlength{\itemsep}{3pt} \settowidth{\labelwidth}{#1.}
     \sloppy
    }}{\end{list}}

\begin{document}
\begin{center}{{\elevenbf
                 $SU(3)$ SKYRMIONS
\footnote{\ninerm\baselineskip=11pt
The work supported by Russian Fund for Fundamental Research, grant
95-02-03868a and by Volkswagenstiftung, FRG}
\\}
\vglue 1.0cm
{\tenrm V. B. KOPELIOVICH, B. E. STERN \\}
{\tenit Institute for Nuclear Research of the Russian Academy of
Sciences,\\ 60th October Anniversary Prospect 7A, Moscow 117312, Russia\\}
\vglue 0.8cm
{\tenrm ABSTRACT}}
\end{center}
\vglue 0.3cm
{\rightskip=3pc
 \leftskip=3pc
 \tenrm\baselineskip=12pt
 \noindent
  The consideration of the bound skyrmions with large strangeness
 content is continued. The connection between 
 $B=2$ $SO(3)$-hedgehog and $SU(2)$-torus is investigated and the
 quantization of the dipole-type configuration with large strangeness
 content is described.
 \vglue 0.6cm}
 \vglue 0.1cm
\baselineskip=14pt
\elevenrm

 1. At the conference Quarks-$94$ we presented preliminary results of 
our studies of $SU(3)$ skyrmions - configurations of
chiral fields which are described by $8$ functions of $3$ variables
\cite{1}. Using a special program allowing for the minimization
of the energy functionals depending on ten functions (with $2$ unitarity
conditions imposed) we obtained the following results.

In the sector with baryon number $B=1$ the $SU(2)$ hedgehog and in 
the sector with $B=2$ the $SU(2)$ torus are local minima in $SU(3)$ 
configuration space. The local rotations in "strange" direction
do not allow to decrease the energy of mentioned static 
configurations which are believed to be absolute minima of static 
energy for $B=1$ and $2$.

The new local minimum was found in the $B=2$ sector with large 
scalar strangeness content $SC$, close to $0.5$. This configuration
was obtained starting with two $B=1$ skyrmions located in $(u,s)$ 
and $(d,s)$  $SU(2)$ subgroups of $SU(3)$ in optimal (attractive)
relative orientation and optimal distance between topological
centers. Applying our minimization algorithm we obtained the 
configuration of the dipole type consisting of two deformed $B=1$
hedgehogs in different $SU(2)$ subgroups with binding energy about
half of that of the torus, i.e. about $0.04$ of the static energy of
$B=1$ hedgehog. 

We investigated the flavor symmetric $(FS)$ case when all meson 
masses in the effective lagrangian are equal to the pion mass. The 
configuration found in \cite{1} possesses remarcable symmetry
properties.

Later we extended our studies to flavor symmetry broken $(FSB)$
case with kaon mass included into the lagrangian.
Here we shall discuss our latest results concerning $FSB$ effects,
the connection between $SO(3)$ hedgehog and $SU(2)$ torus, and
the quantization of $SU(3)$ skyrmions.  \\
 
 2. The parametrization we used for the unitary $SU(3)$ matrix $U$
incorporating $8$ chiral fields is
$$        U = U_L(u,s) U(u,d) U_R(d,s)            \eqno    (1)
$$
where  one  of  SU(2)-matrices,  e.g. $U(u,d)$   depends  on two
parameters:
$$        U(u,d)= \exp(ia \lambda_2 ) \exp(ib \lambda_3 )
\eqno (2) $$
 
The chiral and flavor symmetry breaking terms in the lagrangian density 
can be presented in the following form:

$$ M_{m.t.} = \frac{1}{8} F_{\pi}^2 m_{\pi}^2 (2-v_1-v_2) +
 \frac{1}{8} (2F_{K}^2 m_{K}^2-F_{\pi}^2 m_{\pi}^2) (1-v_3)  \eqno (3) $$

The real parts of diagonal matrix elements of matrix $U$ for this case 
are expressed through the functions parametrizing our ansatz 
according to:

$$ v_1=c_{b} c_{a}f_0 - s_{b} c_{a} f_3  $$
$$ v_2=c_{b}c_{a}q_0 + s_{b}c_{a}q_3 $$
$$ v_3= f_0q_0 - f_3q_3 + s_{a} [s_{b}(f_1q_2-f_2q_1)-
   c_{b}(f_2q_2+f_1q_1)]                       \eqno  (4)     $$

The functions $f_{i}$ and $q_{i}$ parametrize the $(u,s)$ and $(d,s)$
solitons in the following way:

$$ \tilde{U}_L(u,s)= f_0 + i(f_1\tau_1+f_2\tau_2+f_3\tau_3) $$
$$ \tilde{U}_R(d,s)= q_0 + i(q_1\sigma_1+q_2\sigma_2+q_3\sigma_3) \eqno(5) $$

where $\tau_{k}$ and $\sigma_{k}$ are the Pauli matrices of 
$(u,s)$ and $(d,s)$ $SU(2)$ subgroups of $SU(3)$. Scalar strangeness
content of the configuration is defined as
$$ C_S= \frac{1- \bar{v}_3}{3-\bar{v}_1-\bar{v}_2-\bar{v}_3} $$
with averaging/integration over the whole soliton being performed.

Special care was taken to improve the baryon number stabilization
of the configuration, which was especially important when $FSB$
terms were added to the lagrangian.  
The inclusion of the kaon mass term leads to the 
increase of masses of solitons up to $1982 Mev$ and $3895 Mev$ for
$B=1$ hedgehog and $B=2$ dipole - instead of $1702 Mev$ and $3334 Mev$
correspondingly in flavor symmetric case. Here we take 
$F_{K}=F_{\pi}=186 Mev$ and $e=4.12$, as in \cite{1}b. The 
uncertainty
of the calculation does not exceed $\sim 5$ and $\sim 10$ $Mev$ 
for $B=1$ and $2$. 

The other effect of $FSB$ is the reduction of the dimensions of the 
configuration: the kaon cloud which the strange skyrmion consists of
is smaller in dimensions by a factor about $\sim 1.5$ in comparison
with the pion cloud in nonstrange $FS$ case.
In the Fig.1 we show the equal mass density lines for the configurations
of lowest energy for $FSB$ case.

As it was stressed already the configuration we found possesses
remarcable symmetry properties: the functions $a$ and $b$ have symmetry
corresponding to the azimuthal winding $n=2$. In the Fig.2 the function $a$ 
is shown in the plane $z=0$ for $d_0=0.75 Fm$ and for $d_0=0.02$,
also for $FSB$ case. The difference in topological properties of
both configurations is quite clear:  the configuration of more
complicated structure has higher energy.

The physics consequences of the existence of dipole-type configuration
will be considered in Section 4.  \\

 3. We investigated the connection between $SO(3)$ hedgehog and
$SU(2)$ torus, both $B=2$ configurations. Previously we calculated
the masses of both configurations as functions of chiral symmetry
breaking parameter $\mu^2$ \cite{2},\cite{3}. For $\mu =0$ the 
torus is lower in energy by $0.2 \frac{F_{\pi}}{e}$, with 
increasing $\mu$ the difference in masses
decreases and changes sign at $\mu^2 = 0.04  Gev^2$. For realistic
values of chiral symmetry breaking parameter, $\mu^2=8.44 m_{\pi}^2$ 
the mass of $SO(3)$ hedgehog is smaller than the mass of the 
torus by about 
$\sim 30 - 40 Mev$ - depending on the parameters of the model.

A special ansatz allowing for connection between $SO(3)$ hedgehog
and the $(u,d)$ $SU(2)$ torus was proposed by B.Schwesinger. 
It is the following one:

$$U = U_L U_4 U_M U_8 U_4^{\dagger} U_L^{\dagger}  \eqno(6)  $$

$$ U_L = exp(-i\phi \lambda_2) exp(2ig_1 \lambda_3) $$
$$ U_M = exp(-ig_1 \lambda_3) exp(i\chi) exp(ig_1\lambda_3) $$
$$ U_4 = exp(ig_2 \lambda_4) $$
$$ U_8 = exp(-i\rho \lambda_8 /\sqrt{3}) $$
$$ g_1 = \frac{1}{2} arccos(c_{\theta}/c_{g_2}) $$
$$ g_2 = arccos(\sqrt{c_{\theta}^2 + s_{\theta}^2 s_{\gamma}^2}) \eqno(7) $$

Here $\theta = arccos(z/r)$ is the polar angle, $\phi=arctg(y/x)$
is the azimuthal angle in usual coordinates $(x,y,z)$ space.
$c_{\theta}=cos(\theta)$, $c_{\gamma}=cos(\gamma)$. The parameter
$\gamma$ changes in the interval $(0, \pi/2)$.

The $8$ $SU(3)$ rotated Cartan-Maurer currents defined as
$$ U_8U_4^{\dagger}U_L^{\dagger}U^{\dagger}d_iUU_LU_4U_8^{\dagger} = 
iL_{k,i}\lambda_k  \eqno(8) $$

in this case are equal to
$$L_{1,i}= 2c_a[-l_{1,i}(q_2^2+q_3^2)+l_{2,i}(q_1q_2-q_0q_3)]
 +(1+c_a^2)l_{3,i}(q_0q_2+q_1q_3) + m_{1,i}       $$
$$L_{2,i}= 2c_a[l_{1,i}(q_0q_3+q_1q_2)-l_{2,i}(q_1^2+q_3^2)]
 +(1+c_a^2)l_{3,i}(q_2q_3-q_0q_1)] + m_{2,i}      $$
$$L_{3,i}= 2c_a[l_{1,i}(q_1q_3-q_0q_2)+l_{2,i}(q_0q_1+q_2q_3)]
 -(1+c_a^2)l_{3,i}(q_1^2+q_2^2) + m_{3,i}         $$
$$L_{4,i}=-s_a[l_{1,i}q_1+l_{2,i}q_2+c_al_{3,i}(q_3-s_b)]-
 d_ia(q_0-c_b)  $$
$$L_{5,i}= s_a[-l_{1,i}q_2+l_{2,i}q_1+c_al_{3,i}(q_0-c_b)]
 -d_ia(q_3-s_b) $$
$$L_{6,i}= s_a[l_{1,i}(q_3+s_b)+l_{2,i}(q_0-c_b)-c_al_{3,i}q_1
  -d_iaq_2      $$
$$L_{7,i}= s_a[l_{1,i}(q_0-c_b)-l_{2,i}(q_3+s_b)+c_al_{3,i}q_2
   -d_iaq_1     $$
$$L_{8,i}= -d_ib/\sqrt{3}   \eqno(9)  $$

Here functions $q_i$ parametrize the $SU(2)$ matrix $U_{M}$ embedded
into $SU(3)$, similar to expressions $(5)$ above and should be expressed
through functions $g_1$ and $\chi$ according to $(7)$. $U(L)$ is also
$(u,d)$ $SU(2)$ matrix embedded into $SU(3)$. $l_{k,i}$ and $m_{k,i}$
are $SU(2)$ Cartan-Maurer currents connected with $U_L$ and $U_M$,
$k,i=1,2,3$. The expressions $(9)$ should be substituted into the expression
for the static energy of solitons \cite{1}, see also $(11)$ below.

This (local phase) parametrization allows to connect the $SO(3)$
hedgehog and $SU(2)$-torus: for $\gamma =0$ strangeness changing
angle equals to polar angle, $g_2=\theta$, and with special choice
of profiles $\rho$ and $\chi$ we obtain the $SO(3)$ hedgehog.
Further minimization using our algorithm did not lead to any
decrease of the energy and to changes of the configurations.
Therefore, we conclude that the $SO(3)$ hedgehog with $B=2$ is
a local minimum in $SU(3)$ configuration space.

For $\gamma=\pi/2$ we started with energy which is considerably 
greater than energy of the torus. On the start $g_2=0$, but strangeness
content of the configuration is not zero since the function $\rho$ is 
different from zero. The function $g_2$ remains to be $0$ during 
minimization, and $\rho$ decreases and tends to $0$. SC also tends to $0$.
The minimization performed on the work-station at Siegen University 
during several months allowed to get the torus-like configuration 
corresponding to the local minimum in $(u,d)$ $SU(2)$ subgroup of $SU(3)$.

When we started with some intermediate values of $\gamma$ we received
configurations with strong enhancement in the mass density and
baryon number density along $z$-axis. However, big losses of the
$B$-number during minimization did not allow us to come to some firmly
established final configurations. Most probably, these final 
configurations are above both $SO(3)$ hedgehog and $SU(2)$ torus.
The methods we used certainly should be improved to investigate
the complicated ansatz of the type $(6),(7)$. The final mesh effects
are essential for the ansatz of this type, and special care should be
taken to remove these effects. \\

 4. The quantization of zero modes of $SU(3)$ skyrmions 
was invesigated by one of the 
authors \cite{4}, and here we discuss it briefly for completeness.
The quantization of skyrmions was made previously in the following
cases: for $SU(2)$ skyrmions rotated in the $SU(3)$ collective
coordinates space \cite{5,6}, and also for $SO(3)$ skyrmions
\cite{7}. These two cases do not exhaust the possible variety
of $SU(3)$ skyrmions, and, really, the dipole type configuration
with strangeness content close to $0.5$ \cite{1} cannot be quantized 
using the known algorithm.
The generalization of this quantization procedure for arbitrary
$SU(3)$ skyrmions seems to be actual.

Consider first the rotation energy of solitons.
   The kinetic term is:
$$ E^{kin}_{rot} = \frac{ {F_\pi}^2 }{16} (\tilde{\omega}_1^2 + ... 
 + \tilde{\omega}_8^2)
\eqno (10) $$

   The Skyrme term makes the contribution
$$
E^{Sk}_{rot}= {1 \over 8e^2} \Biggl\{
\vec{s}_{12}^2 +
\vec{s}_{23}^2 +
\vec{s}_{31}^2 +
\vec{s}_{45}^2 +
\vec{s}_{67}^2 +
{3 \over 4} \biggl(
\vec{s}_{48}^2 +
\vec{s}_{58}^2 +
\vec{s}_{68}^2 +
\vec{s}_{78}^2 \biggr) + $$
$$ + {1 \over 4} \biggl(
\vec{s}_{46}^2 +
\vec{s}_{47}^2 +
\vec{s}_{56}^2 +
\vec{s}_{57}^2 +
\vec{s}_{14}^2 +
\vec{s}_{15}^2 +
\vec{s}_{16}^2 +
\vec{s}_{17}^2 + 
\vec{s}_{24}^2 +
\vec{s}_{25}^2 +
\vec{s}_{26}^2 +
\vec{s}_{27}^2 +
\vec{s}_{34}^2 +
\vec{s}_{35}^2 +
\vec{s}_{36}^2 +
\vec{s}_{37}^2 \biggr) + $$
$$ + {{\sqrt{3}} \over 2} \biggl(
\vec{s}_{84}
\bigl(
\vec{s}_{16} +
\vec{s}_{34} -
\vec{s}_{27} \bigr) +
\vec{s}_{85}
\bigl(
\vec{s}_{17} +
\vec{s}_{26} +
\vec{s}_{35} \bigr) +
\vec{s}_{86} \bigl(\vec{s}_{14} +\vec{s}_{25} -\vec{s}_{36} \bigr) +
\vec{s}_{87} \bigl(\vec{s}_{15}-\vec{s}_{24}-\vec{s}_{37}\bigr)\biggr)+ $$
$$ + {3 \over 2} \Bigl(
\vec{s}_{12} \bigl(
\vec{s}_{45} +
\vec{s}_{76} \bigr) +
\vec{s}_{23} \bigl(
\vec{s}_{47} +
\vec{s}_{65} \bigr) + 
\vec{s}_{13} \bigl(
\vec{s}_{64} +
\vec{s}_{75} \bigr) +
\vec{s}_{45} \vec{s}_{67} \Bigr) \Biggr\} \eqno (11) $$

with $\vec{s}_{ik}=\tilde{\omega}_i \vec{L}_l - \tilde{\omega}_k
\vec{L}_i $, $i,k=1,2,... 8$.
The $\tilde{\omega}$-s are linear combinations of body-fixed angular
velocities $\omega$-s, see $(12)$ below.
The expression for static energy can be obtained from this one by 
substitution $\vec{s}_{ik}=2[\vec{L}_i\vec{L}_k]$. Note that in
\cite{1} few terms in the expression $E_{Sk}$ proportional to
$\sqrt{3}/2$ are missed by misprint.
The explicit expressions for $L_i$   depend  on  the  ansatz  for
$SU(3)$-matrix $U$.

The moments of inertia of the system ($8$ diagonal and $28$ off-diagonal)
can be calculated from expressions $(10),(11)$ using the connection
between $\tilde{\omega}$-s and body-fixed angular velocities of 
rotation in $SU(3)$-configuration space $\omega$-s:
$$ \tilde{\omega}_i=(R_{ik}(V^{\dagger})-R_{ik}(T))\omega_{k}
 \eqno(12) $$
Real orthogonal matrices $R_{ik}$ are defined as
$$ R_{ik}(V)={1 \over 2} Tr\lambda_i V \lambda_k V^{\dagger} \eqno(13) $$
and $R_{ik}(V^{\dagger}) = R_{ki}(V)$.
$U_0 = VT$, $V=U(u,s)exp(ia\lambda_2)$, $T=exp(ib\lambda_3)U(d,s)$.
The expressions for the moments of inertia are too bulky to be reproduced
here, and we calculated them numerically using $(10),(11)$ and $(12),(13)$
avoiding the explicit analytical expressions for them.
For the strange skyrmion molecule we obtained that there are 4 different
diagonal inertia, $\Theta_1=\Theta_2=\Theta_N$, $\Theta_3$ which is about 
$0.3$ smaller than $\Theta_1$, 
$\Theta_4=\Theta_5=\Theta_6=\Theta_7=\Theta_S$ and
$\Theta_8$ which is a bit greater than $\Theta_4$ \cite{4}. In view
of these symmetry relations the dipole-type biskyrmion can be quantized
like the configuration possessing axial symmetry, at least for the
lowest possible value of angular momentum, $J=0$.

To obtain quantization conditions for skyrmions with arbitrary
strangeness content the Wess-Zumino term in the action
should be investigated \cite{4}.
The quantization condition for $SU(2)$ 
skyrmions located incidently in $(u,d)$ subgroup and quantized 
with $SU(3)$ collective coordinates was  obtained previously in
\cite{6} and has the form
$$ Y_R = N_c B/3      \eqno(14) $$
where $Y_R$ is the so called right hypercharge characterizing the
$SU(3)$ irrep under consideration, $N_c$ is the number of colors 
in the underlying QCD, $N_c=3$ usually, $B$ is the baryon number 
of the system. This quantization condition can be generalized to
$$ Y^{min}_R = {2 \over \sqrt{3}} dL^{WZ}/d\omega_8 = 
N_c B (1 - 3 C_S)/3      \eqno(15)  $$
where $C_S$ is the scalar strangeness content of the soliton \cite{4}.

The formula $(15)$ can be obtained from the Wess-Zumino term
written in the elegant form by Witten \cite{8}.
It is valid exactly
for any soliton obtained from $(u,d)$-$SU(2)$ - solitons by means
of rotation into "strange" direction. In this case $C_S={1 \over 2}
sin^2\nu$, $\nu$ being the angle of rotation. $(15)$ is valid also
for $SO(3)$ solitons considered in \cite{7}. For these solitons
$SC={1 \over 3}$ and $Y_R=0$.
$(15)$ was checked numerically  for the dipole-type configuration
with $C_S=0.5$ found recently in \cite{1}. 
The Wess-Zumino term which defines the quantum numbers of the 
configuration also can be expressed through the currents $L_{k,i}$.
In general case there are $8$ different Wess-Zumino functions according
to the relation

$$L^{WZ}=L^{WZ}_i\omega_i=
 \frac{N_c}{24\pi^2} \int (R_{ik}(U_0)+\delta_{ik})W_k\omega_i d^3x
 \eqno (16)  $$

The summation over coinciding indices is assumed here, $i,k=1,...8$.

The real orthogonal matrix 
$R_{ik}=\frac{1}{2}Tr \lambda_iU_0\lambda_kU_0^+ $. 
The most important for us is function $W_8$
$$W_8= -\sqrt{3}(L_1L_2L_3) + (L_8L_4L_5) + (L_8L_6L_7) \eqno(17) $$

$(L_1L_2L_3)$ is the mixed product of the vectors $\vec{L_1}$,
$\vec{L_2}$, $\vec{L_3}$.
To calculate the functions $W_k$ as well as expressions $(10)$, $(11)$
above we used the expressions for the 
Cartan-Maurer currents of the dipole-type configuration defined as
$TU_0^{\dagger}d_iU_0T^{\dagger}=iL_{k,i}\lambda_k $. 
We obtained \cite{1}:
$$ \begin{array}{ll}
L_{1i}=s_ac_al_{3i}, &
L_{2i}=d_ia,  \nonumber \\[10pt]
L_{3i}=(c_{2a}l_{3i} - r_{3i})/2 + d_ib, & L_{4i} =c_a l_{1i},
 \nonumber \\[10pt]
L_{5i}=c_a l_{2i}, & L_{6i} =s_a l_{1i} + r_{1i}(b), \nonumber \\[10pt]
L_{7i}=s_a l_{2i} + r_{2i}(b), & L_{8i} =  \sqrt{3} (l_{3i}+r_{3i})/2.
\end{array} \eqno (18) $$

   Here $r_1(b)= r_1c_b - r_2s_b$ ,  $r_2(b)= r_1c_b + r_2 s_b$, $l_{ki}$ and
$r_{k,i}$ are left and right $SU(2)$ C-M currents, $k,i=1,2,3$.

In this case the following relation takes place approximately:
$$ Y^{min}_R ={2 \over \sqrt{3}} dL^{WZ}/d\omega_8 = - (B_L+B_R)/2 
  \eqno  (19)  $$
where $B_L$ and $B_R$ are the baryon numbers located in left
$(u,s)$ and right $(d,s)$ $SU(2)$ subgroups of $SU(3)$, or vice
versa. In the case we investigate $B_L=B_R=1$. We obtained 
the relation $(19)$ numerically \cite{4}, but we guess 
that it can be obtained also analytically.

After standard quantization procedure \cite{6},\cite{7} the simplified
expression for the rotation energy of the dipole solitons is
$$E_{rot} = \frac{C_2(SU_3) - 3 Y_R^2/4}{2 \Theta_S} +
\frac{N(N+1)}{2} \bigl({1 \over \Theta_N} - {1 \over \Theta_S}
\bigr) + \frac{3 (Y_R - Y^{min}_R)^2}{8 \Theta_8}  \eqno   (20)  $$

$C_2(SU_3)=(p^2+q^2+pq)/3+p+q$ is the
second order Casimir operator of $SU(3)$ group depending on the numbers
of upper and lower indices $p$ and $q$ in the tensor describing the
$SU(3)$ irrep $(p,q)$.
$N$ is the right isospin equal to the isospin of the isomultiplet
with $Y=Y_R$. 
The difference between $\Theta_N$ and $\Theta_3$ is neglected
here for the estimate as well as few interference moments of inertia
different from zero, e.g. $\Theta_{46}$ and $\Theta_{57}$ which are
at least one order of magnitude smaller than diagonal inertia.
The angular momentum of configurations we discuss has the lowest 
possible value $J=0$.

It is clear that for $\Theta_8 = 0$ it should be $Y_R=Y^{min}_R$,
otherwise this quantum correction will be infinitely large.
With $Y^{min}_R$ given by $(14)$ this is just the Guadagnini's
quantization condition \cite{6}.

For $Y_R=Y^{min}_R=-1$ the following $SU(3)$ irreps for dibaryons
are allowed: octet $(p,q)=(1,1)$, decuplet $(3,0)$ and antidecuplet
$(0,3)$. The corresponding energies including static mass of
the soliton and $E_{rot}$ are equal to $4.44$, $5.0$ and $5.5 Gev$
which should be compared with central values of masses of baryons 
octet and decuplet $2.64$ and $3.05 Gev$.
The $SU(3)$ irreps with
$Y_R$ different from $Y_R^{min}$ are also allowed. Many of these
states should be bound relative to the strong interactions, see
also discussion of the Casimir energy below.\\

The mass splittings within $SU(3)$ multiplets of dibaryons obtained
from strange skyrmion molecule are defined by $FSB$ terms in the 
lagrangian, as usually, and are within $150-200$ $Mev$. 

Note that for the case of quantization of $SU(2)$ bound skyrmion in 
$SU(3)$ collective coordinates space the following multiplets 
appeared with $Y_R=B$, for $B=2$: antidecuplet $(0,3)$, $27$-plet 
$(2,2)$, $35$-plet $(4,1)$, $28$-plet $(6,0)$. All these multiplets 
have nonstrange
components, and the ratio of strangeness to $B$-number changed down
to $-3$ in the $28$-plet \cite{9}. \\

 5. The investigations we performed are not complete because we cannot 
make a statement that all possible $B=2$ 
configurations in $SU(3)$ flavor space have been investigated. 

However, we established connection between $SO(3)$ hedgehog and
$SU(2)$-torus and found that both are local minima in $SU(3)$ 
configuration space. New local minimum with large strangeness
content found recently is investigated also in the flavor symmetry
broken case.

The quantization of this new configuration is performed. The known
previously quantization condition is generalized for the states
with strangeness content different from zero. The lowest $SU(3)$ 
multiplets
which appear in this case do not contain nonstrange components,
they are in some sense more "strange" than multiplets obtained
with quantization condition \cite{6}.

The uncertainty in the values of binding energies of states obtained
by means of quantization of the strange skyrmion molecule \cite{1} 
is the smallest
one in comparison with other states \cite{7}, \cite{9} due to
cancellations of the poor known Casimir energies which control the
absolute values of masses, according to present understanding \cite{10},
\cite{11}. Indeed, the molecule consists of two slightly deformed
unit skyrmions, therefore the Casimir energy of the molelule should be 
close to approximately twice of that of unit hedgehog and should not make
big influence on the binding energies. The results we obtained \cite{4} are in
general qualitative agreement with the results obtained recently in \cite{12}
where the interaction potential of two hyperons was estimated at
relatively large distances.
More detailed calculations are being in progress now.

We are thankful to Bernd Schwesinger who initiated the studies of 
$SU(3)$ skyrmions for many discussions and useful suggestions. The
part of the work concerning the connection between $SO(3)$ soliton and
$SU(3)$ torus was done with his participation.
We are also indebted to G.Holzwarth and H.Walliser for permanent 
interest
in the problems of $SU(3)$ skyrmions and useful discussions. One of
the authors (V.B.K.) thanks V.A.Rubakov for discussions, and
ICTP, where the part of this work was done, for the hospitality.

\vglue 0.4cm
{\elevenbf\noindent References}
\vglue 0.2cm

\vglue 1.5in
{\elevenbf Figure captions}
\vglue 1.0cm
Fig.1 Equal mass density lines for 
the dipole-type configuration with $SC \approx 0.48$, binding 
energy $(\simeq70\pm5)\mbox{MeV}$, $d_0 \simeq 0.75\mbox{Fm}$, 
plane $x=0$, $F_\pi=186\mbox{MeV}$, $e=4.12$ \\

Fig.2 The function $a$ defining the relative local orientation
of deformed $(u,s)$ and $(d,s)$ hedgehogs in isospace.
Plane $z=0$, a) $d_0=0.75\mbox{Fm}$, $M=3.90\mbox{Gev}$ 
b) $d_0=0.02\mbox{Fm}$, $M=4.02\mbox{Gev}$.

\end{document}